\documentclass[12pt,a4]{article}\input{epsf.tex}
\begin{document}
\begin{center}

{\bf Level density and radiative strength functions of dipole
$\gamma$-transitions in $^{163}$Dy}\\
{\bf  V.A. Khitrov,  Li Chol, A.M. Sukhovoj}\\
{\it Frank Laboratory of Neutron Physics, Joint Institute
for Nuclear Physics, 141980, Dubna, Russia }\\ \end{center}

In the report [1] presented at the 11-th Symposium ``Capture Gamma-Ray
Spectroscopy", intensities of the two-step cascades in $^{163}Dy$ measured
in \v{R}e\v{z} were reproduced with involving the idea of "scissors mode"
built on the head-states with different structure of their wave functions.
From the comparison of the experimental data and calculated spectra
(distorted by the random fluctuations of the widths), the authors conclude
about the significant influence of this specific excitations on the cascade
$\gamma$-decay process.

There are some doubts about this result for the following reasons:
$^{163}$Dy is the well deformed nucleus with rather strong quadrupole
deformation. The spectrum of the excited states, structure of their wave
functions and matrix elements of the transitions between them in such nuclei
are well enough described in the framework of quasiparticle-phonon model
of nucleus [2] with taking into account only quasi-particle excitations,
quadrupole and octupole phonons. The notion of significant influence of
the ``scissors mode" on the mentioned nuclear parameters is not used in this
approach. Moreover, analysis [3] of the averaged intensities of cascades
exciting both known ``scissors mode" and states of other type that the
cascade $\gamma$-decay of compound state after the thermal neutron capture
is not selective to within some tens percent. I. e., the known ``scissors
mode" states, quasi-particle, phonon or more complicated states are excited
by cascades practically with equal probability.

According to the notions of the quasiparticle-phonon nuclear model [2],
the probability of the neutron capture is determined by the single-particle
components of the wave function of the neutron state.  As it was estimated
by V.G.Soloviev, their contribution into the total normalization of wave
function of deformed nucleus equals $10^{-9}$. So, the authors of QPNM did
not need to introduce some exotic hypotheses, like that used in [1], for
description of such process as the emission the primary $\gamma$-transition.

\section{New data on the parameters of the cascade $\gamma$-decay}
Analysis [4] of a bulk of the information on the two-step cascades,
made in FLNP without the model notions on level density $\rho$ and radiative
strength functions $k$, showed that these data cannot be reproduced without
the presence of step-like structure in level density and corresponding
deviations of $k$ from the simple model dependences. For the fist time,
the possibility of the step-like structures in the level density and
corresponding thermodynamical characteristics of a nucleus was pointed
out in [5].
\section{Analysis}

Using our early [6] experimental data on cascade $\gamma$-transitions from
the $^{162}$Dy$(n,\gamma)$ reaction we determined the dependence [7] of
the two-step cascade intensity  $I_{\gamma\gamma}(E_1)$ on the primary
transition energy $E_1$, and also level density $\rho$ and radiative strength
functions  $k=\Gamma_{\lambda i}/(E_\gamma^3A^{2/3}D_\lambda)$ which allow us
to reproduce $I_{\gamma\gamma}(E_1)$ (shown in Fig.~1) with zero deviation
from the experiment.

Although efficiency of the spectrometer used in the experiment [6] was not
enough, rather specific form of energy dependence of the cascade intensity
(considerable concentration of  $I_{\gamma\gamma}$ at the energy of their
intermediate levels $E_i \simeq 0.5B_n$) decreases systematical uncertainties
of the procedure [7] for the primary transitions of cascades with the energy
$E_1 \leq 2$ MeV. The most considerable errors of procedure [7] --
decomposition of the experimental spectrum into two components corresponding
to solely primary transitions and to solely secondary transitions -- leads
to re-distribution of cascade intensities in $^{163}Dy$ (Fig.~1) between the
intervals of the primary transition energies 2-3 and 3-4 MeV. Modelling
of influence of this uncertainty on the desired parameters of the cascade
$\gamma$-decay process shows that this does not lead to principle change
in the forms of the energy dependences of level density and radiative
strength functions. For the energy of the primary transition $E_1\leq 2$ MeV
in Fig.~1, corresponding overage in cascade intensity does not exceed 0.1 of
the given value. This conclusion follows from the extrapolation [8] of the
cumulative sums of cascade intensities in some energy intervals of their
intermediate levels to the zero detection threshold of individual cascades.
This systematical error leads, in practice, to insignificant variation in
values of the desired parameters.

Level density and radiative strength functions of $E1$ and $M1$ cascade
transitions, which allow simultaneous reproduction of cascade intensity
 $I_{\gamma\gamma}(E_1)$ (Fig.~1) and the mean value of the total radiative
width $\Gamma_{\lambda}=150$ meV [9] of neutron resonances in $^{162}Dy$,
are shown in Figs.~2 and 3, respectively.
As in the other studied even-odd nuclei, level density in $^{163}Dy$ also
considerably less of [10] in the excitation interval from 1 to 3 MeV.
Theoretical basis for qualitative explanation of such energy dependence
was obtained in [5]. In accordance with the main idea of A.V. Ignatyuk and
Yu.V. Sokolov, qusi-particle level density is the sum of densities of levels
with 1, 3, 5... quasi-particles for even-odd nuclei. In the interval 
between the energies of
breaking of corresponding Cooper pair, level density changes very weakly,
at least for some first broken pairs. And energy of a nucleus, most
probably, is passed for excitation of its vibrations. The only correction
which is necessary to achieve well agreement between the experiment and
calculation [5] is the shift to the higher value of the energy of appearance
of 3 quasi-particles by about 1 MeV.

The increase in radiative strength functions in the interval
$1.5 \simeq E_1  \simeq 4.5$ MeV results, most probably, from the change in
the ratio between the quasi-particle and collective components of wave
functions of the cascade intermediate levels in the region of the biggest
deviation of their density from the predictions of the Fermi-gas model.

\section{On ambiguity of different methods to analyse cascade intensities}

Method [4] does not allow one, even in principle, to get unique values of
the level density and radiative strength functions and in that case when
experimental data do not contain systematical and statistical errors.
Modelling of this situation shows that such their asymptotical uncertainty
for the available experimental data cannot be less than $\approx 20\%$.
This result is to:

(a)~exceeding of the number of the determined parameters over the number
of experimental points, and

(b)~specific form of the functional dependence of $I_{\gamma\gamma}$ on
the desired parameters $\rho$ and $k$.

The possibility to get reliable information on  $\rho$ and $k$ strongly
increases if some hypothesis is tested by means of comparison between the calculated
within according to this hypothesis and experimental distributions of cascade
intensities. In general case, some model ensemble of the $\rho$ and $k$
values allow one to calculate some value $M_{\gamma\gamma}(E_1)$ of cascade
intensity and bilaterally symmetrical to it distribution
$M_{\gamma\gamma}(E_2)$ intensity of cascades to their different final levels.

The case $M_{\gamma\gamma}(E_1)+M_{\gamma\gamma}(E_2)=
 I_{\gamma\gamma}(E_1)+I_{\gamma\gamma}(E_2)$ is the necessary condition
for correspondence between the model and experimental values of $\rho$ and $k$
but it is not enough condition. It is obvious, because the condition
$M_{\gamma\gamma}(E_1)=I_{\gamma\gamma}(E_1)$ does not follow from previous
equation. This means that the experimental spectra can be reproduced with
the help of much larger set of values of $\rho$ and $k$ that it is necessary
for reproduction of the data in Fig.~1.

The authors of [1] and, for example, [11] do not take this important
circumstance into consideration.

This work was supported by RFBR Grant No. 99-02-17863.
\begin{flushleft}
{\large\bf References}
\end{flushleft}
\begin{flushleft}
1. F. Becvar et al.,
   Eleventh International Symposium on Capture Gamma-Ray Spectroscopy\\
~~~~and Related Topics, Pruhonice, September 2-6, 2002, World Scientific,\\
~~~~~Ed. J.Kvasil, P. Cejnar, M.Krticka, p. 726\\
2. V.G. Soloviev, Part Nucl.,  (1972)  {\bf V3(4)} 770\\
3. V.A. Khitrov, A.M. Sukhovoj,
In:  X International Seminar on Interaction
of Neutrons with \\
~~~~Nuclei,  Dubna, May 2002,
E3-2003-10, Dubna, 2003,
p. 156\\
4. E.V. Vasilieva, A.M. Sukhovoj, V.A. Khitrov,
Physics of Atomic Nuclei, (2001) {\bf 64(2)} 153\\
5. A.V. Ignatyuk, Yu.V. Sokolov, Yad. Fiz., (1974)  {\bf 19} 1229\\
6. S.T. Boneva et al., Izv. AN SSSR, Ser. Fiz., (1986) {\bf 50} 1832\\
7. S.T. Boneva et. al., Nucl.  Phys., (1995) {\bf A589} 293\\
8. A.M. Sukhovoj, V.A. Khitrov, Phys. of Atomic Nuclei, (1999) {\bf 62(1)} 19\\
9. S.F. Mughabghab, Neutron Cross Sections. V. 1. Part B. N.Y. Academic
Press, (1984)\\
10. W. Dilg, W. Schantl, H. Vonach and M. Uhl, Nucl. Phys., (1973) {\bf A217}
 269\\
11. F.Becvar et al., Phys.Rev., (1995) {\bf C52} 1278\\
~~~~F.Becvar, P.Cejnar, R.E.Chrien, J.Kopecky, Phys.Rev., (1992) {\bf C46} 1276\\   
~~~~A. Voinov, A.Schiller, M.Guttormsen, J.Rekstad, S.Siem, 
Nucl. Instr. Meth Phys. Res.\\
~~~~ (2003) {\bf A497} 350\\ 
12. S.G. Kadmenskij, V.P. Markushev, V.I. Furman, Sov. J. Nucl. Phys. {\bf 37}
(1983) 165
\end{flushleft}
\begin{figure}[htbp]
\begin{center}
\leavevmode
\epsfxsize=10.5cm
\epsfbox{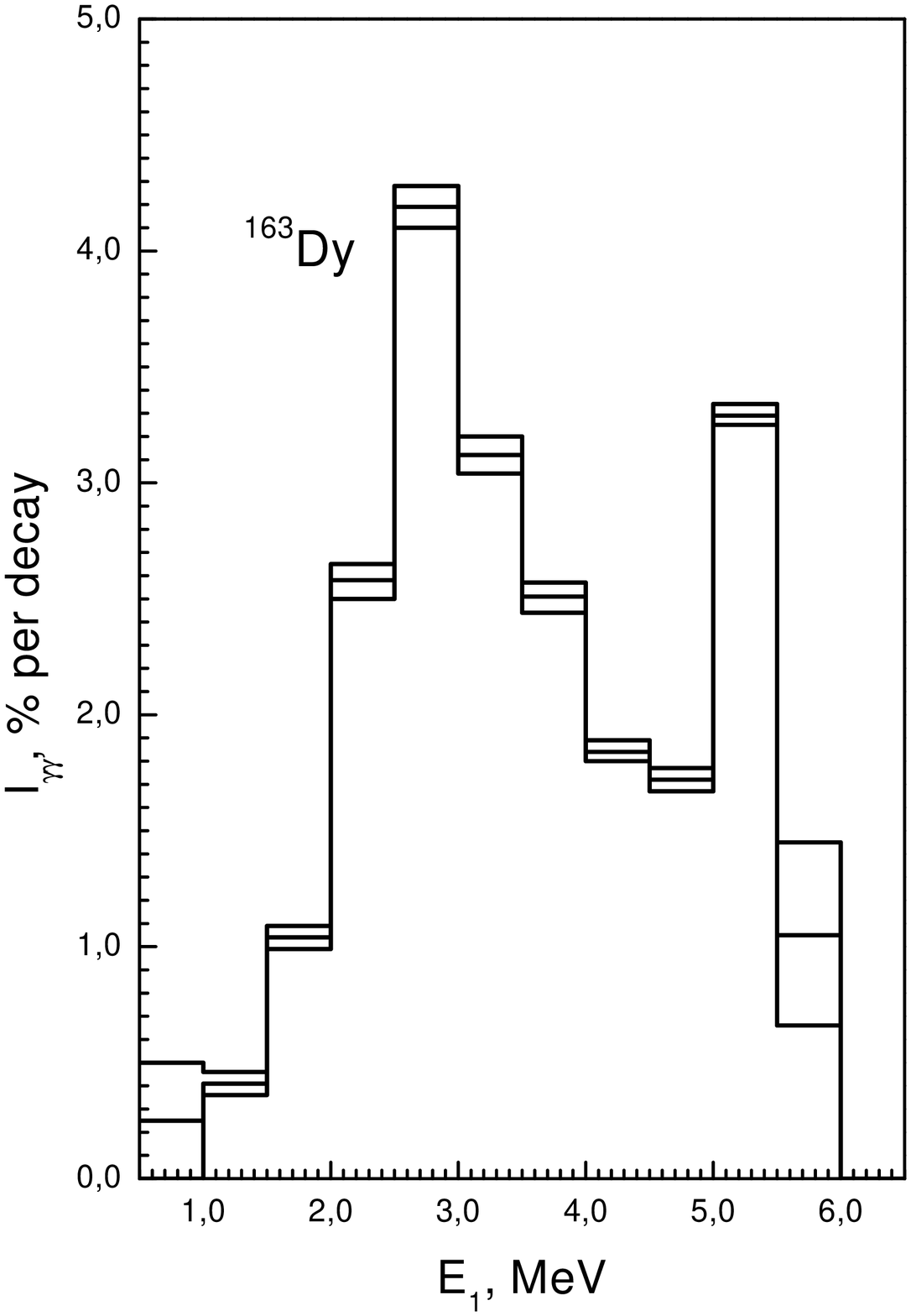}
\end{center}
\hspace{-0.8cm}
{\bf Fig.~1} The total experimental intensities (in \% per decay) of 
two-step cascades (summed in energy bins of 500 keV) with ordinary
statistical errors as a function of the primary transition energy.
\end{figure}

\begin{figure}
\begin{center}
\leavevmode
\epsfxsize=12.5cm
\epsfbox{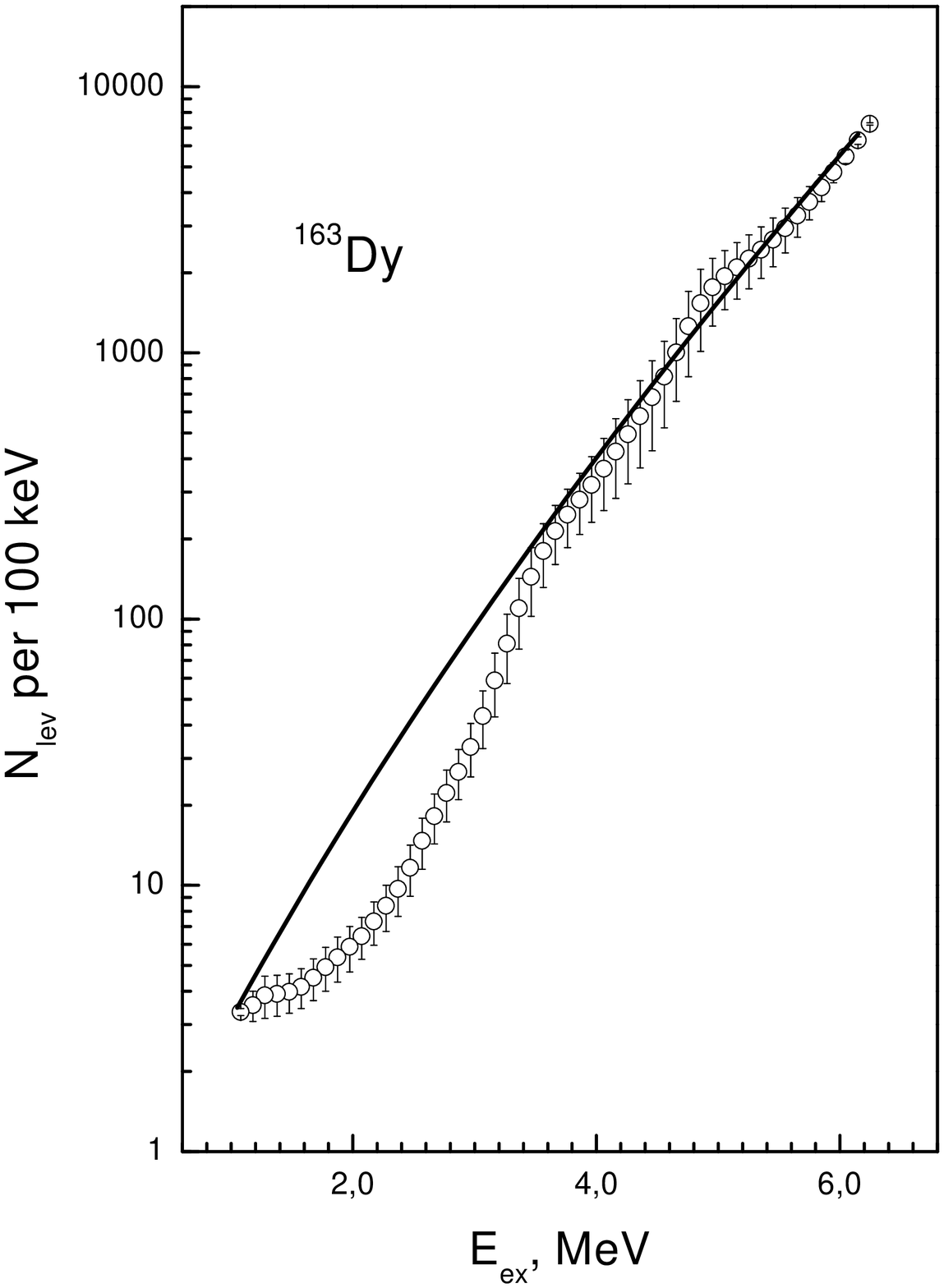}
\end{center}
\hspace{-0.8cm}

{\bf Fig.~2} The interval of probable values of the level density enabling
the reproduction of the experimental cascade intensity and total radiative 
width of capture state. The line represents predictions of model [10].
\end{figure}

\begin{figure}
\begin{center}
\leavevmode
\epsfxsize=12.5cm
\epsfbox{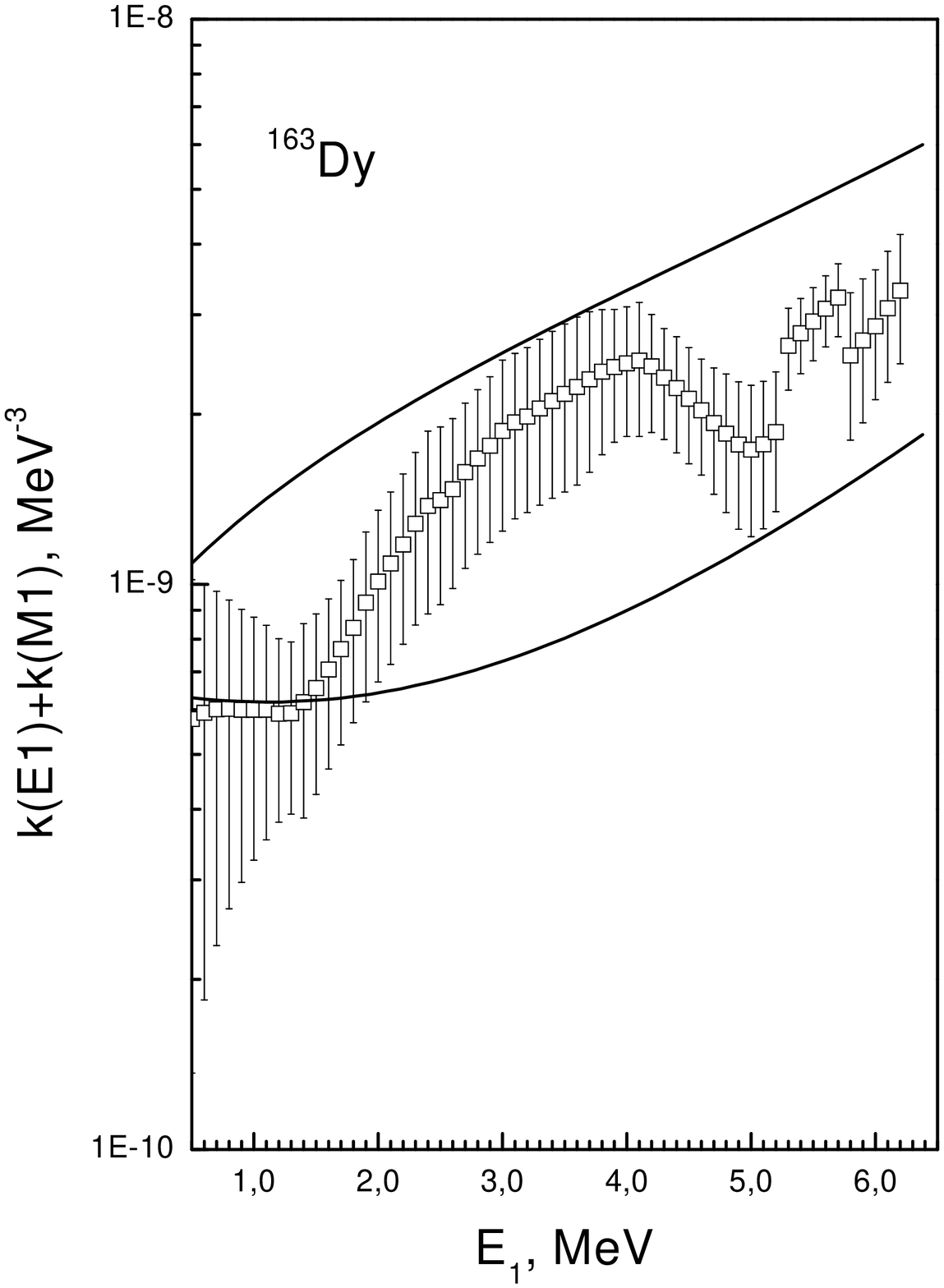}
\end{center}
\hspace{-0.8cm}
{\bf Fig.~3} The probable interval of the sum strength function $k(E1)+k(M1)$
(points with error bars) providing the reproduction of the experimental data.
The upper and lower curves represent the extrapolation of the GEDR ``tail" 
into the region below $B_n$ and the predictions of model [12], respectively.\end{figure}

\end{document}